# The overlap of neural selectivity between faces and words: evidences from the N170 adaptation effect

Xiao-hua Cao [1], Chao Li [1], Carl M Gaspar [2], Bei Jiang [1]

1. Department of Psychology, Zhejiang Normal University, Jinhua 321004, China;

2. Center for Cognition and Brain Disorders, Hangzhou Normal University, Hangzhou 310000, China.

Authors: Xiao-hua Cao, caoxh@zjnu.cn

Chao Li, lichao87_2006@163.com

Gaspar M Carl, earl.leatherman@yahoo.co.uk

Bei Jiang, bluestone710@163.com

**Abstract:** Faces and words both evoke an N170, a strong electrophysiological response that is often used as a marker for the early stages of expert pattern perception. We examine the relationship of neural selectivity between faces and words by using a novel application of cross-category adaptation to the N170. We report a strong asymmetry between N170 adaptation induced by faces and by words. This is the first electrophysiological result showing that neural selectivity to faces encompasses neural selectivity to words, and suggests that the N170 response to faces constitutes a neural marker for versatile representations of familiar visual patterns.

**Key words**: N170; adaptation; visual expertise; neural selectivity

**Introduction**

The human brain has limited resources for representing the visual world [2]. It makes sense that different categories of images share their visual representations to some degree. Many non-face expert stimuli recruited the face-selective parts of the visual system [7-10,18]. Two of the most familiar types of visual patterns are faces and written language. Despite their difference in appearance, current studies suggested that there may be a strong relationship between the visual representations for faces and words. For example, reading ability enhanced the left fusiform activation which induced a small competition with faces at this location [3]. Recently, it was revealed that that faces and words both evoke an N170, a strong electrophysiological response that is often used as a marker for the early stages of expert pattern perception[1,14,17].These evidences suggested that both faces and words do evoke a representation that is shared to a certain degree. However, thus far, nothing is known about the nature of overlap between electrophysiological selectivity to faces and words. Behavior studies revealed that, although the configural information critical for identity processing is different between faces and words, there are many similarities of the two categories , for example, extensive and long-term exposure, canonical upright orientation, high level expertise, predominantly individual-identity level processing, and featural information critical for identity processing, which may contribute to shape the expert pattern perception[15]. Moreover，another study found that the expertise increased the functional overlap between face and car[16].　Thus, these evidences above may be implicated that the neural selectivity of early face processing is overlap, even encompasses that of words. For the first time, we used the cross-category adaptation paradigm [13](face adaptor followed by word test and vice versa) to directly reveal the relation of the early expert perception between faces and words.

## Methods

### Participants

Sixteen subjects (7 males) were recruited from local universities and paid for their participation (age range: 19-29 years, mean 21.8 years). All subjects were right handed with normal or corrected-to-normal vision. Informed consent was obtained from all subjects, and the study was approved by the ethical committee of Zhejiang Normal University.

### Stimuli

Grayscale pictures of faces, Chinese characters and houses were used in this experiment. Faces were images of 72 individuals (36 male and 36 female), displaying neutral facial expression. Using Adobe Photoshop, these faces were cropped to remove external features (hair, ears, and jaw line) and replaced with the same oval contour used by Eimer et al. (2010)[5]. All 72 Chinese characters used in this study had a left-right configuration, high frequency values (from 0.01372 to 0.62457) (the modern Chinese frequency dictionary, 1986), with the number of the strokes varying from 7 to 14, and were presented in Song font. In addition, 72 grayscale images of houses were used which were similar with Eimer et al (2010) [5]. The face stimuli were 180 × 276 pixels, subtending an angle of 4.0° × 6.2° from a viewing distance of 90 cm. The character and house stimuli were 198 × 198 pixels, subtending an angle of 4.5° × 4.5° from a viewing distance of 90 cm. All stimuli had the same luminance contrast.

**Procedure**

The participants were asked to sit on a chair, with a distance of 90 cm away from the 17″ CRT monitor (1024 × 768 pixel resolution), on which all stimuli were presented against a dark grey background. E-Prime software was used for stimulus presentation and behavioral response collection (Psychology Software Tools, Pitts-burgh, PA).

Subjects were tested in a dimly lit room. In each trial, an adaptor stimulus and a test stimulus were presented sequentially for 200 ms each with a 200 ms inter-stimulus interval and followed by a 1500 ms intertrial interval, which was consistent with Eimer et al (2010) [5]. Adaptor stimuli and test stimuli were always different. One of three possible adaptor stimuli - Faces (F), Chinese characters (C), or houses (H) - was followed by one of two possible target stimuli (F or C). Therefore, there were 6 conditions: FF, CF, HF, CC, FC, HC. Equal numbers of each category were presented in random order in each of the four experimental blocks. There were 108 trials in each block, 12 of which were target trials with a red outline shape aligned with the outer contours of the stimulus shape. These target trials were randomly intermixed with the experimental trials and presented with equal probability as adaptor stimuli or test stimuli. Participants were instructed to press a response button following the second picture presentation when they detected a target. Response buttons were counterbalanced across subjects.

**EEG Recording and Data Analysis**

EEG was recorded using a 128-channel HydroCel Geodesic Sensor Net, with an electrode placed on the Vertex (Cz) serving as reference for the online recording. Electrode impedances were kept below 50 kΩ. Signals were digitized at a 500 Hz sampling rate, and amplified with a 0.1–200 Hz elliptical bandpass filter. EEG data were offline digitally filtered with a 0.3–30 Hz band-pass filter and epoched from 200 ms before to 800 ms after stimuli onset with a 100 ms pre-stimulus baseline. Trials with artifacts exceeding ±100 μV were rejected. Any subject with more than 30% bad segments would be excluded from the group-average. The remaining EEG data were re-referenced to the average of channels.

EEG data were analyzed for nontarget. trials only. A group of the larger response channels over the left hemisphere (58, 64, 65) and right hemisphere (90, 95, 96) were analyzed. EEG waveforms were averaged separately for each presentation condition of adaptor or test stimuli. Based on visual inspection of the individual data, the N170 time-window was defined as 130-210 ms for adaptor stimuli and 140-220 ms for test stimuli. Repeated-measures analyses of variance (ANOVA) were performed on the peak amplitudes and latencies of N170 component with stimuli category and recording hemisphere serving as within-subject factors.

## Results

### Adaptor stimuli data

The results of adaptor stimuli were shown in Figure 1, Table 1 and Table 2. A two-way repeated-measures ANOVA of N170 peak amplitudes and latencies was conducted for adaptor categories (faces, characters, houses) and hemispheres (left hemisphere, right hemisphere). A main effect of adaptor category on N170 amplitude ($F(2, 30) = 14.619$, $p < 0.001$, $\eta_p^2=0.494$) and an interaction between adaptor categories and hemispheres ($F(2, 30) = 8.227$, $p = 0.003$, $\eta_p^2=0.354$) were found. Post hoc tests revealed that the N170 amplitudes elicited by faces were much larger than Chinese characters ($p = 0.001$) and houses ($p < 0.001$). There was a main effect of adaptor category on N170 latency ($F(2, 30) = 10.990$, $p = 0.001$, $\eta_p^2=0.423$) and an interaction between adaptor and hemisphere (F (2, 30) = 6.678, p = 0.005, $\eta_p^2$=0.308). Paired comparisons showed that the N170 was earlier for characters relative to faces (*p* = 0.003) and houses (*p* = 0.001).

**(Insert Fig 2 about here)**

**(Insert Table 1 about here)**

**(Insert Table 2 about here)**

**Test stimuli data**

We analyzed the N170 amplitude elicited by the test stimuli in six combinations of adaptor and test category. One of three possible adaptor stimuli - Faces (F), Chinese characters (C), or houses (H) - was followed by one of two possible target stimuli (F or C). Therefore, there were 6 conditions: FF, CF, HF, CC, FC, HC. For example, CF refers to the N170 amplitude to faces in condition CF, after adapting to a character. The average wave shapes and the topographical map are shown in Figure 2.

The results of the 6 conditions analysis were shown in Table 3 and Table 4. The repeated-measures ANOVA of N170 peak amplitudes and latencies were conducted for the 6 conditions (CC, CF, FC, FF, HC, HF) and hemispheres (left hemisphere, right hemisphere). For the N170 peak amplitude, a main effect of test category (F(5, 75) = 12.910, p < 0.001, $\eta_p^2$=0.463) and an interaction between conditions and hemisphere (F(5, 75) = 6.877, p = 0.001, $\eta_p^2$= 0.314) were found. Post hoc tests are shown in

Table 5. There was a main effect of 6 conditions on N170 latency ($F_{(5, 75)} = 7.295$, $p < 0.001$, $\eta_p^2 = 0.327$).

To test the relationship between face N170 adaptation and word N170 adaptation, we focused the analyses as followed:

The adaptation effect of both face and character-specific N170s are typically measured relative to a house-induced adaptation effect: HF is significantly larger than FF in both the left ($p < 0.001$) and right hemisphere ($p < 0.001$), The results replicated the previous studies [4,5] HC is significantly larger than CC in both the left ($p = 0.001$) and right hemisphere ($p < 0.001$).

The results of the cross-category adaptation effect between faces and Chinese characters were revealed by the following analysis:

We analyzed the face-specific N170 (FN) adaptation effect in the FF, CF and HF conditions. The amplitude in the FF condition was significantly smaller than in the CF (left, $p = 0.001$; right, $p < 0.001$) and HF (left, $p < 0.001$; right, $p < 0.001$) conditions. And the amplitude in the CF was similar to that in the HF condition. It suggested that Chinese characters and houses have the same effect to the rapid followed face stimuli, namely, compare with face, both the Chinese characters and houses almost could not produce the adaptation effect to the faces. The results of the adaptation effect of the Chinese character N170 (CN) in the CC, FC and HC conditions are as follows: The amplitude of CN in the

HC condition was significantly larger than those in the FC (left, $p = 0.012$; right, $p < 0.001$) and CC (left, $p = 0.001$; right, $p < 0.001$) conditions. Most importantly, the amplitude of CN was similar in the FC and CC condition. The results suggested that both faces and Chinese characters produced the similar CN adaptation effect to the following Chinese characters.

In summary, the dramatic results show that faces produce an N170 adaptation effect on Chinese characters, but Chinese characters cannot elicit a similar adaptation effect on faces.

**(Insert Fig 2 about here)**

**(Insert Table 3 about here)**

**(Insert Table 4 about here)**

**(Insert Table 5 about here)**

**Discussion**

The purpose of this study was to reveal the relationship of the neural selectivity between faces and words. In this experiment, our results show an asymmetric N170 adaptation effect between faces and Chinese characters, that is, faces produced full adaptation to Chinese characters, not vice versa. This surprising phenomenon provides many insights into the neural selectivity of the N170 component.

Together with the principle of adaptation [6,11], our results allow us to examine the nature of the overlap in neural selectivity to faces and Chinese characters. Just as shown in Figure 3, the first three scenarios imply the symmetric adaptation, while the last two refer to the asymmetric one. The fourth scenario (Figure 3-D) shows that the neural selectivity of CN completely contains the neural selectivity of CN. But previous studies suggested that N170 elicited by face stimuli was related to holistic /configural processing [12], which is a critical property of face perception but not of Chinese character [15]. Moreover, if this scenario is true, based on the principle of adaptation [6,11], the adaptation effect of Chinese characters on faces would be stronger than that of faces to Chinese characters. So this possibility can't be supported. Therefore, the only possibility is that the neural selectivity of FN completely encompasses that of CN, as is shown in Figure 3-E, which can reasonably explain our main results and coincide with our hypothesis. This possibility also suggested that the complex component with bigger neural selectivity (e.g. FN) could produce complete adaptation of the smaller one (e.g. CN).

**(Insert Fig 3 about here)**

The above analysis demonstrates that the face N170 reflects a broad range of functions that includes those involved in Chinese character processing. Future behavioral and brain functional studies may reveal the nature of the differences in information processing that underlie the overlap in neural selectivity between faces and Chinese characters.

**Table 1.** Mean peak amplitudes and latencies (standard deviations) of the N170 elicited by the adaptor stimuli at left hemisphere (LH) and right hemisphere (RH).

|  | Amplitudes (µV) | | Latencies (ms) | |
| --- | --- | --- | --- | --- |
|  | LH | RH | LH | RH |
| Character | -6.12 (3.15) | -6.01 (3.71) | 156.52 (8.20) | 151.94 (9.55) |
| Face | -7.07 (3.06) | -8.35 (4.53) | 161.06 (10.76) | 160.77 (9.15) |
| House | -5.40 (3.11) | -6.27 (4.58) | 159.58 (4.29) | 160.42 (7.40) |

**Table 2.** Statistical results of a 3 categories × 2 hemispheres repeated-measures ANOVA on the peak amplitudes and latencies of the N170 component elicited by adaptor stimuli.

|  | Amplitudes | | Latencies | |
| --- | --- | --- | --- | --- |
|  | F | $p$ | F | $p$ |
| Category | (2, 30) = 14.619 | 0.000* | (2, 30) = 10.990 | 0.001* |
| Hemisphere | (1, 15) = 0.655 | 0.431 | (1, 15) = 0.426 | 0.524 |
| Category × Hemisphere | (2, 30) = 8.227 | 0.003* | (2, 30) = 6.678 | 0.005* |

* Indicates p-values lower than 0.05, which was considered as significant.

**Table 3.** Mean peak amplitudes and latencies (standard deviations) of the N170 elicited by the test stimuli as a function of adaptor stimuli at LH and RH.

|    | Amplitudes (µV) | | Latencies (ms) | |
| --- | --- | --- | --- | --- |
|    | LH | RH | LH | RH |
| CC | -5.43 (2.85) | -5.33 (2.51) | 169.46 (11.70) | 166.75 (14.17) |
| FF | -4.66 (2.30) | -5.94 (2.45) | 181.67 (17.31) | 178.33 (12.21) |
| FC | -5.82 (3.14) | -5.83 (2.15) | 171.17 (12.49) | 171.46 (15.43) |
| CF | -7.20 (2.83) | -9.43 (3.97) | 175.58 (10.80) | 176.25 (9.88) |
| HC | -7.24 (3.00) | -7.65 (2.77) | 181.08 (16.62) | 179.08 (13.15) |
| HF | -7.06 (2.42) | -8.36 (3.18) | 175.42 (10.37) | 177.75 (11.68) |

**Table 4.** Statistical results of a 6 categories × 2 hemispheres repeated-measures ANOVA on the peak amplitudes and latencies of the N170 component elicited by test stimuli.

|    | Amplitudes | | Latencies | |
| --- | --- | --- | --- | --- |
|    | F | $p$ | F | $p$ |
| Category | $(5, 75) = 12.910$ | 0.000* | $(5, 75) = 7.295$ | 0.000* |
| Hemisphere | $(1, 15) = 1.470$ | 0.244 | $(1, 15) = 0.291$ | 0.597 |
| Category × Hemisphere | $(5, 75) = 6.877$ | 0.001* | $(5, 75) = 0.794$ | 0.513 |

* Indicates p-values lower than 0.05, which was considered as significant.

**Table 5.** Paired comparisons for the different-adaptation conditions of the N170 amplitudes elicited by the test stimuli.

|    |    | Left hemisphere | | Right hemisphere | |
|----|----|-----------------|-------|------------------|-------|
|    |    | Mean difference | Sig   | Mean difference  | Sig   |
| CC | HC | 1.809           | 0.001 | 2.324            | 0.000 |
| CC | FC | 0.385           | 0.443 | 0.497            | 0.092 |
| FC | HC | 1.424           | 0.012 | 1.827            | 0.000 |
| FF | HF | 2.397           | 0.000 | 2.418            | 0.000 |
| FF | CF | -2.538          | 0.001 | -3.486           | 0.000 |
| CF | HF | -0.141          | 0.747 | -1.068           | 0.054 |

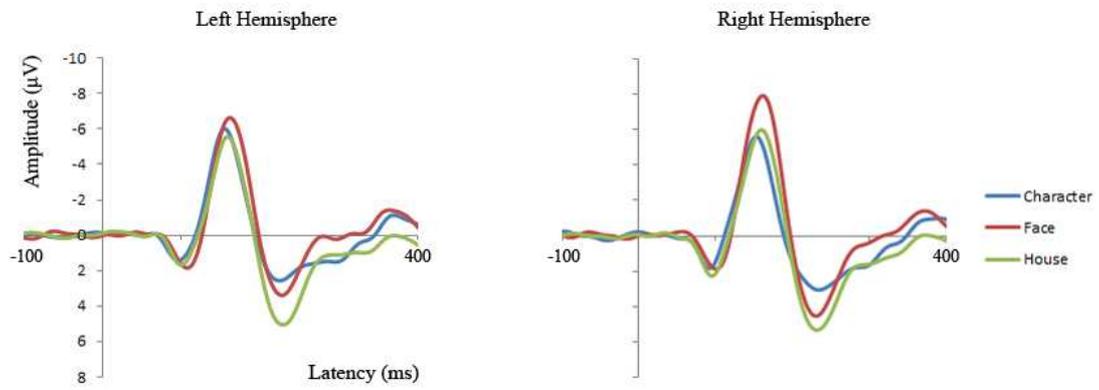

**Figure 1.** Grand averaged N170 waveforms elicited by adaptor stimuli (Character, Face and House) at left and right hemispheres separately.

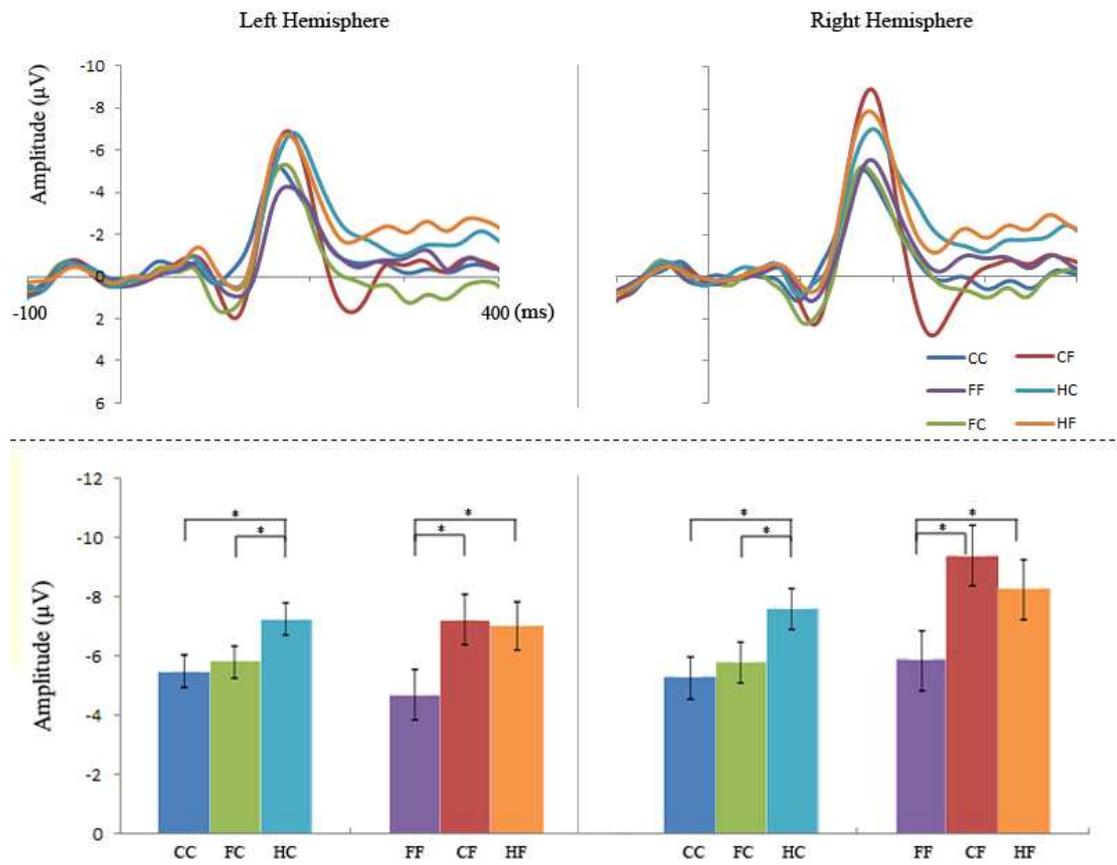

**Figure 2.** Top row: The averaged N170 waveforms elicited by test stimuli at the left and right hemispheres separately. Bottom row: The N170 amplitudes elicited by the test stimuli in different-adaptation conditions.

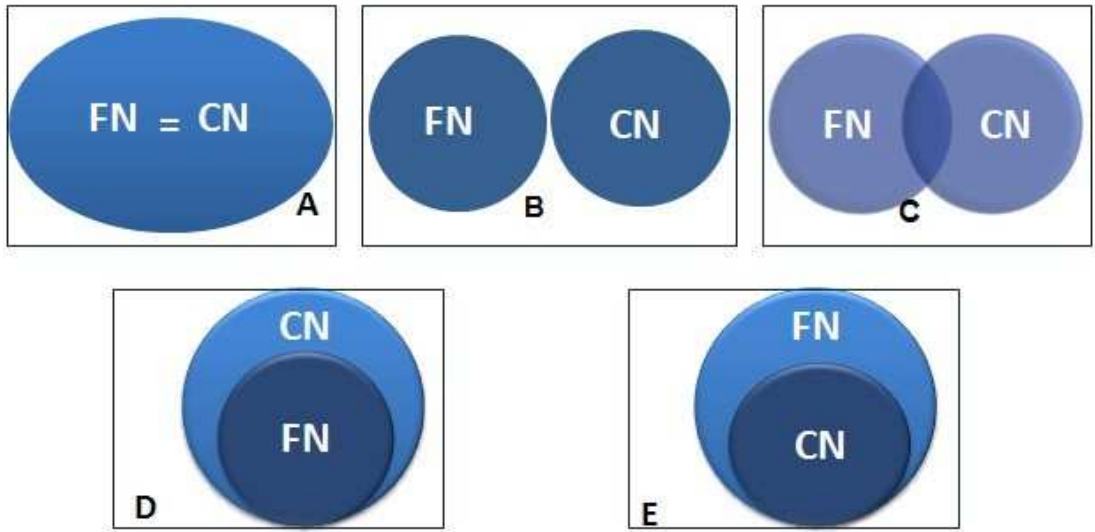

**Figure 3.** The five possibilities of the relations between face-specific N170 (FN) and Chinese character-specific N170 (CN)